\documentclass[11pt]{article}
\usepackage{fullpage}
\usepackage{amsmath, amssymb, amsthm}
\usepackage{hyperref}

\newtheorem{theorem}{Theorem}

\newtheorem{corollary}[theorem]{Corollary}
\newtheorem{definition}[theorem]{Definition}

\newcommand{\NP}{\mathsf{NP}}
\newcommand{\coNP}{\mathsf{coNP}}
\newcommand{\ZPP}{\mathsf{ZPP}}
\newcommand{\TFNP}{\mathsf{TFNP}}

\newcommand{\Poly}{\mathsf{P}}
\newcommand{\SsubTwoP}{\mathsf{S}_2^\Poly}

\title{Search versus Decision for $\SsubTwoP$}
\author{Lance Fortnow\\Illinois Institute of Technology}
\date{}

\begin{document}

\maketitle

\begin{abstract}
We compare the complexity of the search and decision problems for the complexity
class $\SsubTwoP$. While Cai (2007) showed that the decision problem is contained in $\ZPP^{\NP}$, we show that the search problem is equivalent to $\TFNP^{\NP}$, the class
of total search problems verifiable in polynomial time with an $\NP$ oracle.
This highlights a significant contrast: if search reduces
to decision for $\SsubTwoP$, then $\Sigma_2^\Poly \cap \Pi_2^\Poly \subseteq
\ZPP^{\NP}$.
\end{abstract}

\section{Introduction}

The complexity class $\SsubTwoP$, introduced by Canetti \cite{Canetti96} and
Russell and Sundaram \cite{RussellS98}, is defined by a game between two
competing provers. A language $L$ is in $\SsubTwoP$ if there exists a
polynomial-time predicate $P$ such that for $x \in L$, there is a strategy for
the first prover that wins against any move of the second prover, and for $x
\notin L$, there is a strategy for the second prover that wins against any move
of the first. Alternatively we can think of an exponential-sized matrix generated
by $P$ where $x$ is in $L$ if there is a row of all ones and $x$ is not in $L$
if there is a column of all zeros.

Cai~\cite{Cai01} showed that the decision problem for $\SsubTwoP$ is in
$\ZPP^{\NP}$, suggesting it is relatively easy, at least compared to
$\Sigma_2^\Poly\cap\Pi_2^\Poly$. However Cai's algorithm does not necessarily 
find a witness, the row of all ones or the column of all zeros. Cai's algorithm might eliminate all possible columns guaranteeing
that $x$ is in $L$ but leave many possible rows. In this paper, we explore the 
complexity of this search problem and contrast it
with the complexity of the decision problem.

The search problem appears to be much harder. We
relate this search problem to $\TFNP$, the class of total function problems
introduced by Megiddo and Papadimitriou \cite{MegiddoP91}. Specifically, we
consider $\TFNP^{\NP}$, the class of total search problems where the validity of
a solution can be checked in polynomial time given access to an $\NP$ oracle.

Our main result is that finding an $\SsubTwoP$ witness is complete for
$\TFNP^{\NP}$. While $\SsubTwoP$ is unlikely to contain $\Sigma_2^\Poly\cap\Pi_2^\Poly$, the latter class reduces to the search problem for $\SsubTwoP$.

\section{Preliminaries}
 
\subsection{Definitions}

\begin{definition}[$\SsubTwoP$]
A language $L$ is in $\SsubTwoP$ if there exists a polynomial-time predicate $P$
and a polynomial $q$ such that:
\begin{itemize}
    \item If $x \in L$, then $\exists y \in \{0,1\}^{q(|x|)} \, \forall z \in
    \{0,1\}^{q(|x|)} \, P(x, y, z) = 1$.
    \item If $x \notin L$, then $\exists z \in \{0,1\}^{q(|x|)} \, \forall y \in
    \{0,1\}^{q(|x|)} \, P(x, y, z) = 0$.
\end{itemize}
\end{definition}

\begin{definition}[$\SsubTwoP$-Search]
Given an input $x$ and the predicate $P$ defining an $\SsubTwoP$ language,
output $y$ such that $\forall z \, P(x, y, z) = 1$ or output $z$ such that
$\forall y \, P(x, y, z) = 0$.
\end{definition}

\begin{definition}[$\TFNP^{\NP}$]
A binary relation $R(x, y)$ is in $\TFNP^{\NP}$ if:
\begin{itemize}
    \item $R$ is decidable in polynomial time with an $\NP$ oracle (i.e., $R \in
    \Poly^{\NP}$).
    \item For every $x$, there exists a $y$ such that $R(x, y)$ holds.
    \item The length of $y$ is bounded by a polynomial in the length of $x$.
\end{itemize}
The search problem is: Given $x$, find $y$ such that $R(x, y)$.
\end{definition}

\section{The Equivalence}

\begin{theorem}
$\SsubTwoP$-Search is equivalent to $\TFNP^{\NP}$.
\end{theorem}

\subsection{$\SsubTwoP$-Search in $\TFNP^{\NP}$}
To show that $\SsubTwoP$-Search is in $\TFNP^{\NP}$, we need to show that the
problem is total and that verifying a solution can be done in $\Poly^{\NP}$.
By the definition of $\SsubTwoP$, for every $x$, either there exists a $y$ such
that $\forall z P(x, y, z)$ holds or there exists a $z$ such that $\forall y \neg P(x,
y, z)$ holds. Thus, a witness always exists.
To verify a claimed witness $y$ (asserting $x \in L$), we need to check if
$\forall z P(x, y, z)$. This is a $\coNP$ question, which can be answered by an
$\NP$ oracle. Similarly, verifying a witness $z$ (asserting $x \notin L$)
requires checking $\forall y \neg P(x, y, z)$, which is also a $\coNP$ question.

Therefore, $\SsubTwoP$-Search is in $\TFNP^{\NP}$.

\subsection{$\TFNP^{\NP}$ reduces to $\SsubTwoP$-Search}

We now show that any problem in $\TFNP^{\NP}$ can be reduced to finding an
$\SsubTwoP$ witness. Let $R$ be a total relation in $\Poly^{\NP}$. We construct
an $\SsubTwoP$ predicate $Q$ such that finding a strategy for $Q$ yields a $y$
satisfying $R(x, y)$.

Since $R \in \Poly^{\NP}$, there is a polynomial-time machine $M$ with oracle
access to SAT that decides $R(x, y)$. Let $x$ be an input. We define the
predicate $Q(x, Y, Z)$ as follows.

The first player's move $Y$ consists of:
\begin{itemize}
    \item A candidate solution $y$.
    \item A transcript of oracle queries and answers $\vec{a} = (a_1, \dots,
    a_k)$ for the computation of $M(x, y)$.
    \item For every query $i$ where $a_i = 1$ (asserting the query formula
    $\phi_i$ is satisfiable), a witness $w_i$ satisfying $\phi_i$.
\end{itemize}

The second player's move $Z$ consists of:
\begin{itemize}
    \item An index $j \in \{1, \dots, k\}$.
    \item A witness $w'_j$.
\end{itemize}

The predicate $Q(x, Y, Z)$ evaluates to 1 if all of the following hold:
\begin{enumerate}
    \item The simulation of $M(x, y)$ using oracle answers $\vec{a}$ results in
    acceptance.
    \item For all $i$ where $a_i = 1$, $w_i$ is a valid satisfying assignment
    for $\phi_i$.
    \item If $a_j = 0$ (asserting $\phi_j$ is unsatisfiable), then $w'_j$ does
    \emph{not} satisfy $\phi_j$.
\end{enumerate}

If a valid $y$ exists (which is true by totality), Player 1 can choose $y$, the
correct oracle answers, and valid witnesses for all satisfiable queries. In this
case, conditions 1 and 2 are met. For condition 3, since the oracle answers are
correct, if $a_j = 0$, then $\phi_j$ is truly unsatisfiable, so no $w'_j$ can
satisfy it. Thus, Player 1 has a winning strategy.

Conversely, if Player 1 has a winning strategy $Y$, condition 1 implies $M$
accepts $y$ given $\vec{a}$. Condition 2 ensures all YES answers are correct.
Condition 3 ensures that Player 2 cannot produce a witness for any query
answered NO, implying all NO answers are correct (since Player 2 could play any
witness). Thus, $y$ must be a valid solution to $R(x, y)$.

\section{Implications and Discussion}

This result bridges the gap between the symmetric hierarchy and total search
problems.

\begin{corollary}
If search reduces to decision for $\SsubTwoP$, then $\Sigma_2^\Poly \cap
\Pi_2^\Poly \subseteq \ZPP^{\NP}$.
\end{corollary}

\begin{proof}
Cai \cite{Cai01} showed that $\SsubTwoP \subseteq \ZPP^{\NP}$. If search reduces
to decision for $\SsubTwoP$, then finding an $\SsubTwoP$ witness can be done in
$\ZPP^{\NP}$.

Consider any language $L \in \Sigma_2^\Poly \cap \Pi_2^\Poly$. Since $L \in
\Sigma_2^\Poly$, there is a polynomial-time predicate $A$ such that $x \in L
\iff \exists y \forall z A(x,y,z)$. Since $L \in \Pi_2^\Poly$, there is a
polynomial-time predicate $B$ such that $x \notin L \iff \exists y' \forall z'
B(x,y',z')$. The search problem for $L$ is to find either a $y$ such that
$\forall z A(x,y,z)$ or a $y'$ such that $\forall z' B(x,y',z')$. Since one of
these must exist, the problem is total. Furthermore, verifying a solution
requires checking a universal quantifier, which can be done with an $\NP$
oracle. Thus, the search problem for $L$ is in $\TFNP^{\NP}$.

Since $\TFNP^{\NP}$ is equivalent to $\SsubTwoP$-Search, this implies that
finding witnesses for languages in $\Sigma_2^\Poly \cap \Pi_2^\Poly$ can be done
in $\ZPP^{\NP}$. Consequently, $\Sigma_2^\Poly \cap \Pi_2^\Poly \subseteq
\ZPP^{\NP}$.
\end{proof}

This highlights a significant difference between search and decision for
$\SsubTwoP$. While the decision version is relatively ``easy'' (in
$\ZPP^{\NP}$), the search version captures the full complexity of $\TFNP^{\NP}$.

\section{Conclusion}
We have shown that $\SsubTwoP$ is the ``correct'' complexity class for
characterizing $\TFNP^{\NP}$. This equivalence provides new insights into both
the structure of the symmetric hierarchy and the nature of total search problems
with oracle verification.

\section*{Acknowledgment}
While the results are fully due to the author, this paper was mostly generated using
the large language model Gemini 3 Pro with prompting from the author. The author
takes full responsibility for its contents.

\end{document}